\documentclass[a4paper]{article}

\usepackage{INTERSPEECH2022}
\usepackage{hyperref}
\usepackage{subcaption}
\usepackage{enumerate}

\title{Neural Predictor for Black-Box Adversarial Attacks on Speech Recognition}
\name{Marie Biolkov\'a$^1$\sthanks{\,\, This work was conducted while interning at Sony.},\, Bac Nguyen$^2$}
\address{
  $^1$\'Ecole Polytechnique F\'ed\'erale de Lausanne, Switzerland\\
  $^2$Sony Europe B.V. R\&D Center, Stuttgart Laboratory 1, Germany}
\email{marie.biolkova@epfl.ch, Bac.NguyenCong@sony.com}

\begin{document}

\maketitle
\begin{abstract}
Recent works have revealed the vulnerability of automatic speech recognition (ASR) models to adversarial examples (AEs), \textit{i.e.}, small perturbations that cause an error in the transcription of the audio signal. Studying audio adversarial attacks is therefore the first step towards robust ASR. Despite the significant progress made in attacking audio examples, the black-box attack remains challenging because only the hard-label information of transcriptions is provided. Due to this limited information, existing black-box methods often require an excessive number of queries to attack a single audio example. In this paper, we introduce NP-Attack, a neural predictor-based method, which progressively evolves the search towards a small adversarial perturbation. Given a perturbation direction, our neural predictor directly estimates the smallest perturbation that causes a mistranscription. In particular, it enables NP-Attack to accurately learn promising perturbation directions via gradient-based optimization.  Experimental results show that NP-Attack achieves competitive results with other state-of-the-art black-box adversarial attacks while requiring a significantly smaller number of queries. The code of NP-Attack is available online\footnote{Code available at \url{https://github.com/mariegold/NP-Attack/}}. 
\end{abstract}

\let\thefootnote\relax\footnotetext{Submitted to INTERSPEECH 2022.}

\vspace{-3mm}
\section{Introduction}

There has been significant progress in improving the performance of automatic speech recognition (ASR) based on deep neural networks during the last few years~\cite{graves2013speech,amodei2016deep,hannun2014deep}. As a result, it enables speech recognition technology in many real-world applications, such as Amazon Transcribe, IBM Speech to Text, and Google Cloud Speech to Text. A typical pipeline of an ASR system consists in extracting acoustic features from the audio signal, \textit{e.g.,} the frequency spectrum or mel-frequency cepstral coefficients (MFCCs), then employing an acoustic model that predicts which phonetic units are present. Finally, a language model is used to determine the most likely word sequence. More recently, end-to-end ASR models, which directly output characters or words from audio, have gained popularity. These models have been shown to achieve state-of-the-art performance~\cite{hannun2014deep,watanabe2017hybrid,li2020developing} by replacing the engineering process with the learning process and optimizing the whole network in a single objective function.

Despite their exceptional performance, many studies have revealed that neural networks are vulnerable to adversarial examples (AEs)~\cite{szegedy2013intriguing,carlini2018audio,neekhara2019universal,kong2020adversarial}. These examples are carefully constructed by adding imperceptible perturbations to the inputs, which cause the model to output a specific phrase (\textit{i.e.,} targeted attack)  or any incorrect transcript (\textit{i.e.,} untargeted attack). Most existing attacks on ASR assume the white-box setting~\cite{carlini2018audio,neekhara2019universal}, where the adversary has full knowledge of the model, including the network architecture and parameters. Under this setting, adversarial perturbations are often found via gradient-based optimization such as  the fast gradient sign method~\cite{goodfellow2014explaining} or DeepFool~\cite{moosavi2016deepfool}. However, commercial ASR systems are typically not open source and such information is rarely exposed to the adversary. A more practical attack treats the target ASR model as a black box, \textit{i.e.,} the adversary may only observe the transcribed text~\cite{chen2020devil}. This is much more challenging due to the discrete nature of the output, which does not directly allow gradient computation or estimation. In addition, speech is typically sampled at 16~kHz, making the perturbation search space very high-dimensional even for short utterances. Dimensionality reduction techniques can be applied to make the problem more tractable~\cite{shukla2019black}. Yet, finding a suitable low-dimensional space is not always trivial. One would have to design an efficient search algorithm or resort to combinatorics~\cite{moon2019parsimonious} to tackle this problem. 

Another important constraint in black-box attacks is the maximum number of queries allowed during the optimization, also referred to as query budget. It is clearly unrealistic to have unlimited bandwidth access for querying the target ASR model.  In real-world scenarios, one can query the target model for labeling but cannot exceed the budget or have access to any internal information of the ASR model. This has motivated growing research interest in adversarial attacks with query-limited context~\cite{ilyas2018prior, cheng2018query,cheng2019sign}. However, existing black-box attack methods often require a huge number of queries due to the lack of information about the target ASR model. Consequently, these methods become less applicable with small query budgets. 

In this paper, we demonstrate the feasibility of black-box adversarial attacks in speech recognition by designing a simple and query-efficient method. Our method identifies a small subset of promising perturbation directions through the guidance of a neural predictor. To the best of our knowledge, this is the first method based on neural predictors for black-box attacks. The contributions of this paper are summarized as follows.
\begin{enumerate}[(i)]
    \item We propose NP-Attack, a novel method to generate audio AEs for a black-box ASR system. The idea is to design a neural predictor that estimates the distance to the decision boundary which causes a mistranscription of the audio signal. Unlike other methods relying on substitute models, this neural predictor can be trained with much fewer data since it does not estimate the transcript outputs of the target ASR model.
    \item We conduct several experiments on the LibriSpeech dataset~\cite{librispeech}, where AEs are created to attack the transformer-based ASR model from SpeechBrain~\cite{speechbrain}. Without exposing any prior knowledge about the ASR model, NP-Attack can achieve a better success rate with significantly fewer queries compared to other state-of-the-art black-box methods. 
\end{enumerate}

\section{Related work}
Although AEs have been extensively studied in the image domain, there has been considerably less work dedicated in the speech domain. One of the reasons is because humans are more sensitive to auditory perturbations than visual perturbations~\cite{schonherr2019adversarial,qin2019imperceptible}. We review below some relevant adversarial attack methods in the audio domain.

In particular, Carlini and Wagner~\cite{carlini2018audio} demonstrated the feasibility of audio adversarial attacks on DeepSpeech~\cite{hannun2014deep}, an open-source end-to-end ASR model, with a 100\% success rate. Using the full knowledge of the ASR model parameters, the authors proposed a gradient-based optimization method that minimized the Connectionist Temporal Classification (CTC) loss~\cite{graves2006connectionist}. Adversarial perturbations were updated by backpropagating the gradients through the network and the MFCC layer. As an extension, Sch{\"o}nherr et al.~\cite{schonherr2019adversarial} improved the perceptibility of the adversarial perturbations based on psychoacoustic hiding. Interestingly, Neekhara et al.~\cite{neekhara2019universal} demonstrated the existence of universal adversarial perturbations that are transferable across models with different architectures.

Although white-box audio adversarial attacks showed very promising results, they are quite restricted.  To get a more realistic scenario, recent developments in AEs have shifted to target the black-box scenario. Many approaches relied  on conventional black-box optimization techniques such as evolutionary optimization~\cite{khare2019adversarial} or Bayesian optimization~\cite{shukla2019black}. To simplify the problem, some works~\cite{taori2019targeted,wang2020towards} assumed the knowledge of the log-probabilities given by the target ASR model, which are typically not available to the adversary in the hard-label black-box attack. A common drawback of these methods is that they require many queries to attack a target audio example. Based on the transferability assumption of AEs, Chen et al.~\cite{chen2020devil} introduced a black-box attack method on commercial ASR systems by approximating the target ASR model with a substitute model, which was used to craft AEs. However, unlike in image classification,  more sophisticated techniques are often required to train the substitute model in speech recognition as ASR models usually consist of complicated architectures, including preprocessing, an acoustic model, and a language model. Furthermore, attacks are limited to the most frequently used phrases to make the substitute model more reliable.

\section{Proposed method}
In this section, we formally formulate the problem of finding AEs as an optimization problem. Then, we introduce NP-Attack, a query-efficient approach to solve this problem. Finally, the network architecture and implementation details of NP-Attack are described.

\subsection{Problem formulation}
Consider a trained black-box ASR model as a function $f$ that maps an input audio $\mathbf{x}\in [-1,1]^D$ to a transcript $\mathbf{t} = f(\mathbf{x})$, a sequence of characters or words. Our goal is to find an imperceptible perturbation $\bm{\delta} \in \mathbb{R}^D$ such that the ASR model mistranscribes the input audio signal. Finding such an adversarial perturbation can be formulated as an optimization problem\footnote{To keep the perturbed audio valid, we often perform a clipping operation to $[-1, 1]$. For simplicity, we assume this is included in the ASR model $f$ and do not write it explicitly in the formulation.},
\begin{equation}
\begin{aligned}
\min_{\bm{\delta}} \|\bm{\delta} \|_p  & \quad \text{s.t.,}\quad   f(\mathbf{x}+\bm{\delta}) \neq \mathbf{t} \,,
\end{aligned} \label{eq:form}
\end{equation}
where $\|.\|_p$ is the $\ell_p$-norm indicating a perceptibility metric. Following previous work on audio attacks \cite{carlini2018audio, neekhara2019universal} and to quantify the overall loudness of the perturbation, we will consider the $\ell_{\infty}$-norm  for the remainder of this paper. Unfortunately, a direct optimization to find the minimum-norm perturbation of problem~(\ref{eq:form}) is intractable due to the lack of knowledge about the function $f$. Additionally, the introduction of the max function in the $\ell_{\infty}$-norm makes the problem even harder. 

\begin{figure}[t]
    \centering
    \includegraphics[width=0.65\linewidth]{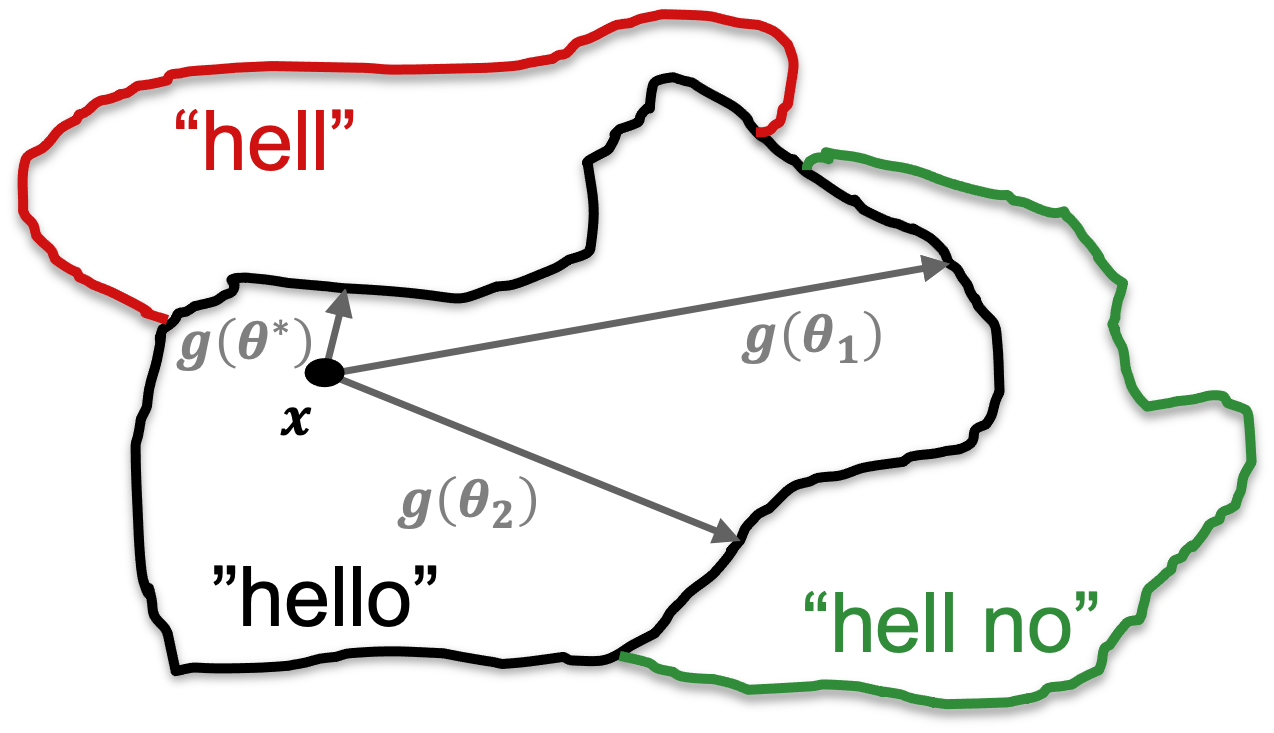}
    \vspace{-0.1cm}
    \caption{An illustration of the refined problem formulation. The input space is divided into several regions corresponding to different transcriptions produced by the target ASR model.}
    \label{fig:boundary}
    \vspace{-0.5cm}
\end{figure}

To overcome these challenges, we follow the boundary-based attack formulation established by Cheng et al.~\cite{cheng2018query}. In particular, the perturbation $\bm{\delta}$ is factorized into a direction vector $\bm{\theta} \in \mathbb{R}^D$ and a magnitude scalar $\lambda \in \mathbb{R}^{+}$, \textit{i.e.}, $\bm{\delta}=\lambda\, \bm{\theta}/\|\bm{\theta}\|$. Given a perturbation direction vector $\bm{\theta}$, the distance from $\mathbf{x}$ to the nearest AE along $\bm{\theta}$ is defined as
\begin{align}
g(\bm{\theta}) = \min_{\lambda > 0} \lambda & \quad \text{s.t.,} \quad  f\left(\mathbf{x}+\lambda \frac{\bm{\theta}}{||\bm{\theta}||}\right) \neq \mathbf{t}\,. \label{eq:g}
\end{align}
Note that $g(\bm{\theta})$ also corresponds to the distance to the decision boundary along $\bm{\theta}$. Using the above definition, problem~(\ref{eq:form}) can be rewritten as
\begin{equation}
\min_{\bm{\theta}} g(\bm{\theta})\,. \label{eq:form2}
\end{equation}
There are several advantages with this formulation. First, it has been shown that the above objective function is locally smooth and continuous~\cite{cheng2018query}. That is, a small change of $\bm{\theta}$ leads to a small change of $g(\bm{\theta})$ (see Fig.~\ref{fig:boundary} for an illustration). Second, instead of searching for the constrained perturbation $\bm{\delta}$, we simplify the problem to searching for a direction vector $\bm{\theta}$, which is an unconstrained optimization. Although computing $g(\bm{\theta})$ in Eq.~(\ref{eq:g}) corresponds to solving another constrained optimization problem with respect to $\lambda$, it requires only a single degree of freedom, making the problem simpler. Interestingly, $g(\bm{\theta})$ can be approximated up to certain accuracy by a two-step search procedure~\cite{cheng2018query}. As a first step, a coarse-grained search is applied to find the range  of magnitudes in which the perturbation causes a mistranscription. More specifically, let $\alpha>0$ be the step size, the coarse-grained  search is done by querying a sequence of points $\{\mathbf{x} + \alpha \,\bm{\theta}/\|\bm{\theta}\|, \mathbf{x} + 2\alpha \,\bm{\theta}/\|\bm{\theta}\|, \dots\}$ one by one until an AE is found, \textit{i.e.}, $f(\mathbf{x} + i\alpha \,\bm{\theta}/\|\bm{\theta}\|) \ne \mathbf{t}$ for some $i>0$. In the second step, we employ a binary search procedure to find the smallest magnitude $\lambda^{*}$ within the range of $[(i-1)\alpha, i\alpha]$ such that $f(\mathbf{x} + \lambda^{*} \,\bm{\theta}/\|\bm{\theta}\|) \ne \mathbf{t}$.

\subsection{Neural predictor}
\begin{figure}[t]
  \centering
 \includegraphics[width=\linewidth]{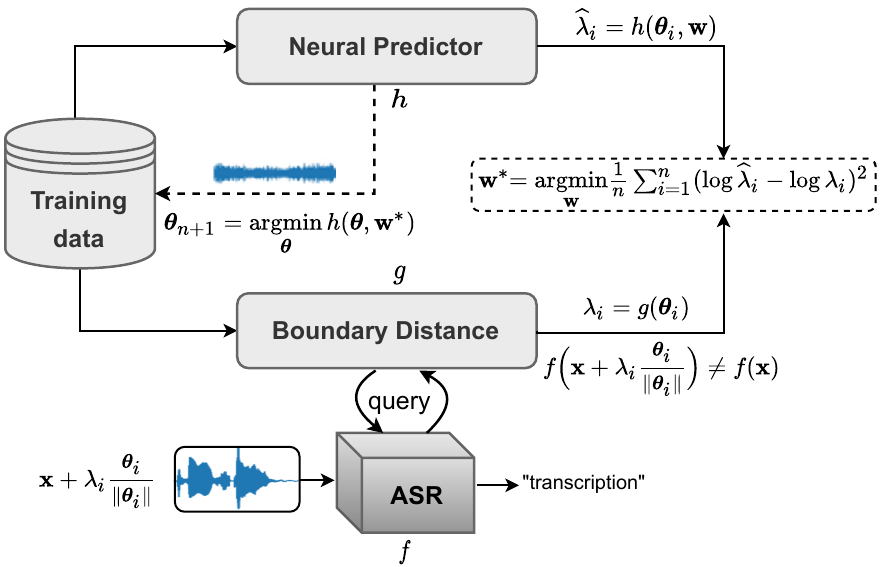}
    \vspace{-0.5cm}
  \caption{Overview of the NP-Attack method.}
  \label{fig:nn}
  \vspace{-0.5cm}
\end{figure}

We aim to solve problem~(\ref{eq:form2}) by progressively fitting a neural predictor as a proxy that estimates the distance from $\mathbf{x}$ to the decision boundary along a given perturbation direction. In the first step, we generate a dataset by querying the target ASR model, then train a neural predictor based on this dataset. In the second step, we use the trained neural predictor to identify a list of promising perturbation directions.  Our neural predictor can substantially accelerate the search process since we have full knowledge of the predictor parameters. The neural predictor is retrained every time a new batch of samples is obtained by querying the target ASR model for the ground-truth distances.

More specifically, we start by generating $n$ training examples $\mathcal{D} = \left\{\left(\bm{\theta}_1, \lambda_1\right), \ldots, \left(\bm{\theta}_n, \lambda_n\right)\right\} \subset \mathbb{R}^D \times \mathbb{R}^{+}$ by querying the ASR model. The ground truth distance from $\mathbf{x}$ to the decision boundary $\lambda_i=g(\bm{\theta}_i)$ is determined for each perturbation direction $\bm{\theta}_i$ via the two-step search procedure as explained in the previous subsection. After constructing the dataset, it is used to train the neural predictor $h(\cdot, \mathbf{w})\colon \mathbb{R}^D \to R^{+}$, parameterized by $\mathbf{w}$. We aim to estimate the distance to the decision boundary $\widehat{\lambda} = h(\bm{\theta}, \mathbf{w})$ by solving the following problem:
\vspace{-1.5mm}
\begin{equation}
    \mathbf{w^*} = \operatorname*{argmin}_{\mathbf{w}} \frac{1}{n}         \sum_{i=1}^n\Big(\log{h(\bm{\theta}_i, \mathbf{w}) - \log{\lambda}_i}\Big)^2.
    \label{eqn:loss}
    \vspace{-2mm}
\end{equation}
To find the next promising perturbation direction, we freeze the trained parameters $\mathbf{w^*}$ and find the next candidate by solving 
\vspace{-1.5mm}
\begin{equation}
    \bm{\theta}_{n+1} = \operatorname*{argmin}_{\bm{\theta}} h\left(\bm{\theta}, \mathbf{w^*}\right)\,. \label{eq:theta}
    \vspace{-2mm}
\end{equation}
Assuming that $h(\bm{\theta},\mathbf{w})$ is differentiable with respect to both $\bm{\theta}$ and $\mathbf{w}$, problems~(\ref{eqn:loss}) and (\ref{eq:theta}) can be solved via gradient-based optimization. Subsequently, we compute the ground-truth distance $\lambda_{n+1}=g(\bm{\theta}_{n + 1})$ by querying the ASR model and add this perturbation direction to the training set $\mathcal{D} := \mathcal{D} \cup (\bm{\theta}_{n+1}, \lambda_{n+1})$.  Next, the predictor parameters are then unfrozen again and modified according to Eq.~(\ref{eqn:loss}) to account for the newly added examples. This process is repeated until the query limit is reached or a solution within the perturbation budget is found. Figure~\ref{fig:nn} illustrates the training process of NP-Attack.  In the beginning, the neural predictor might produce noisy outputs.  To alleviate this problem, we generate a batch of several promising perturbation directions with different random initializations. After progressively adding more examples, the neural predictor is able to produce more reliable outputs. 

\subsection{Network architecture}

\begin{figure}[t]
    \centering
    \includegraphics[width=0.95\linewidth]{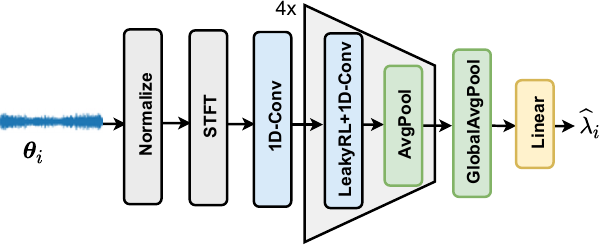}
    \caption{Network architecture of the neural predictor.}
    \label{fig:predictor}
    \vspace{-0.5cm}
\end{figure}

The overall architecture of our neural predictor is shown in Fig.~\ref{fig:predictor}. It maps the perturbation direction $\bm{\theta}$ to a positive scalar value $\widehat{\lambda}$, indicating the distance to the decision boundary along this direction. More specifically, the input is first normalized to have a unit $\ell_{\infty}$-norm. To reduce the temporal dimension of the input, we perform the short-term Fourier  transform (STFT) where the FFT, window, and hop size are set to 1024, 1024, and 256, respectively. Each frequency is then considered as a channel and a 1D convolution is applied to compress the input down to 32 channels. The resulting signal is passed through 4 blocks to further reduce the temporal dimension.  Each block consists of the Leaky ReLU activation with a negative slope of 0.2, 1D convolution, and average pooling with a kernel size of 2. We use a kernel size of 3 for all convolutions and employ weight normalization~\cite{salimans2016weight} for all layers. As the last step, global average pooling is used to remove temporal dimensions of the input, followed by a linear layer to produce the prediction. To ensure the network output is positive, we use the exponential function as an activation function after the linear layer. Note that our neural predictor can be used for any input of arbitrary length.

To train the weights $\mathbf{w}$ of the predictor, we employ the Adam optimizer~\cite{adam} with a learning rate of $10^{-4}$ and an exponential scheduler (a decay rate of 0.99). We use a batch size of 32 and train the model for 300 epochs. The perturbation directions $\bm{\theta}$ are optimized by minimizing the predicted distance to the decision boundary $h(\bm{\theta}, \mathbf{w})$. Each search starts from a random initialization.

\section{Experiments}

\subsection{Experimental setup}

\begin{figure*}[t]
    \begin{subfigure}{.33\textwidth}
      \centering
      \includegraphics[width=\linewidth]{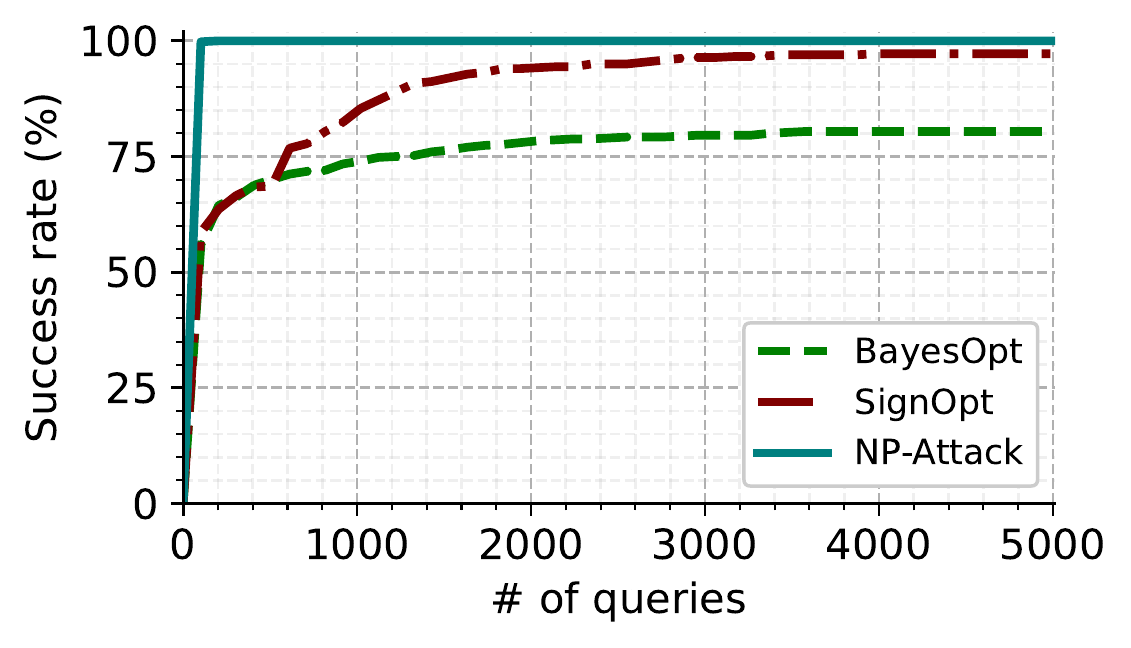}
      \vspace{-0.5cm}
      \caption{$\lambda_{\text{max}} = 0.1$}
      \label{fig:sub-first}
    \end{subfigure}
    \begin{subfigure}{.33\textwidth}
      \centering
      \includegraphics[width=\linewidth]{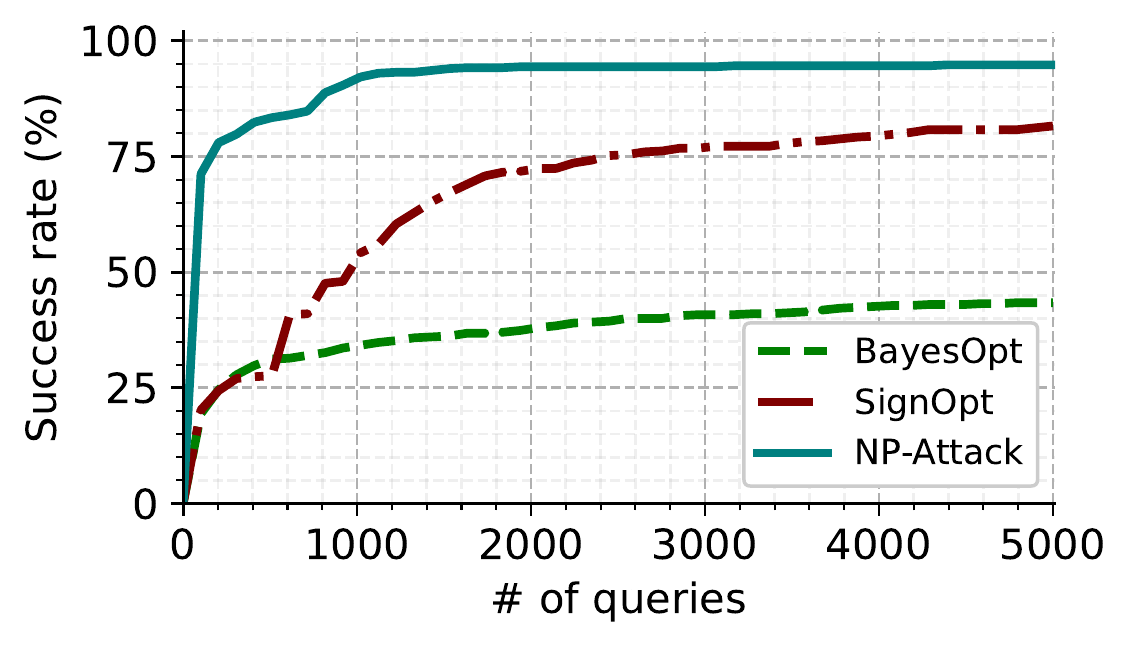}  
      \vspace{-0.5cm}
      \caption{$\lambda_{\text{max}} = 0.05$}
      \label{fig:sub-second}
    \end{subfigure}
    \begin{subfigure}{.33\textwidth}
      \centering
      \includegraphics[width=\linewidth]{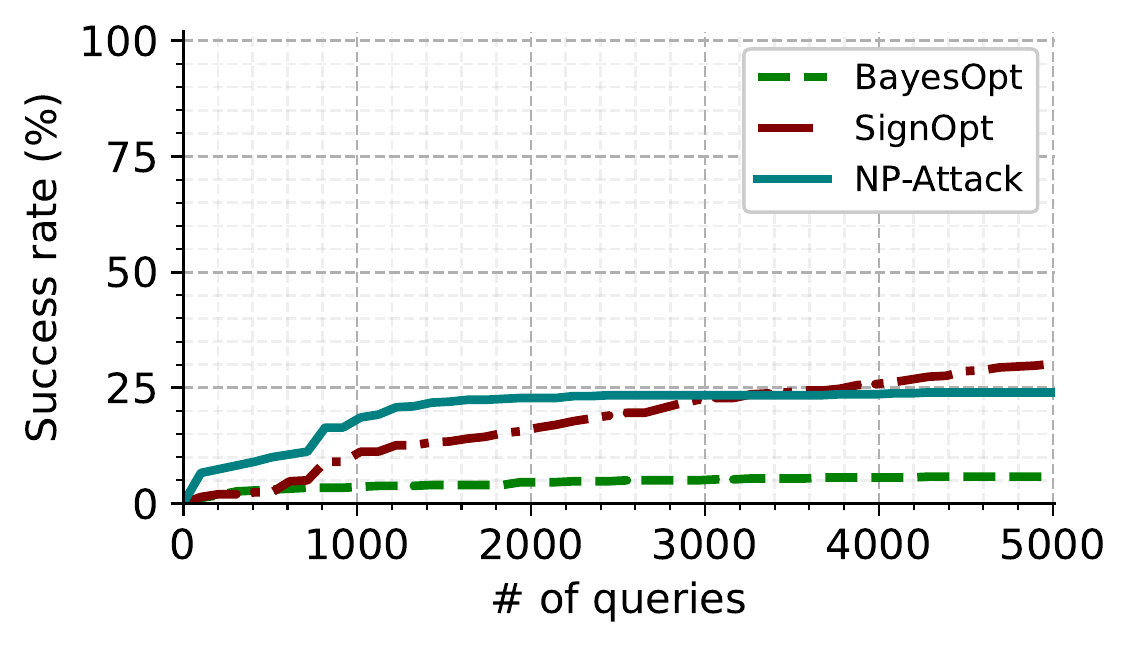}  
      \vspace{-0.5cm}
      \caption{$\lambda_{\text{max}} = 0.01$}
      \label{fig:sub-third}
    \end{subfigure}
    \vspace{-0.1cm}
    \caption{Average attack success rate vs the number of queries for different perturbation budgets.}
    \label{fig:success}
    \vspace{-0.4cm}
\end{figure*}

\textbf{Dataset.} To evaluate the effectiveness of an attack method, we construct a dataset by randomly choosing 100 examples from the LibriSpeech clean test data~\cite{librispeech}. These audio examples are derived from English audiobooks sampled at 16 kHz with the transcript lengths varying from 5 to 10 words. We ensure that all examples are correctly transcribed by the target ASR model.

\textbf{Evaluation metrics.} To measure the performance of an ASR system, we compute the standard word error rate $\text{WER} = (S+I+D)/N_w$, where $S$, $I$, and $D$ indicate the number of substitutions, insertions, and deletions of words, respectively, and $N_w$ is the total number of words in the original phrase. Under a query budget, an attack is considered as successful if the perturbation $\bm{\delta}$ satisfies that $f(\mathbf{x}) \neq f(\mathbf{x}+\bm{\delta})$ and $\|\bm{\delta}\| \le \lambda_{\text{max}}$, where $\lambda_{\text{max}}$ is a perturbation budget. The success rate of an attack method is calculated as $N_{s} / N_{t} \times 100\%$,
where $N_s$ and $N_t$ are the number of successful attacks and the total number of test examples, respectively. In addition, we report the $\ell_{\infty}$-norm of perturbations $\bm{\delta}$, which is a commonly used measure in previous literature of audio attacks. To better account for the energy of the original audio example, we also provide the signal-to-noise ratio (SNR), defined as 
$\mathrm{SNR}(\bm{\delta}) = 20 \log_{10}\left({\frac{||\mathbf{x}||_{\infty}}{||\bm{\delta}||_{\infty}}}\right)$.

\textbf{ASR model.} We target an end-to-end transformer-based ASR model from SpeechBrain~\cite{speechbrain} trained on the LibriSpeech corpus~\cite{librispeech}. The pretrained ASR model achieves a WER of 2.46\% on the clean test set and is publicly available on HuggingFace\footnote{The model can be downloaded from \url{https://huggingface.co/speechbrain/asr-transformer-transformerlm-librispeech}}. Essentially, the model transforms the waveform into a mel-spectrum, then employs an acoustic model comprising of convolutional blocks and a transformer encoder~\cite{NIPS2017_3f5ee243}. The decoding is done using a transformer followed by a beam search. Despite the model being open-source, we do not use any internal information to achieve the attack. Note that NP-Attack can be employed to attack any black-box ASR systems.

\textbf{Threat models.} We compare the performance of NP-Attack with other state-of-the-art black-box hard-label attack methods, including BayesOpt~\cite{Ru2020BayesOpt} and SignOpt~\cite{cheng2019sign}. These methods have proven highly successful in finding image AEs under a low-budget setting. In particular, BayesOpt needs adaptation  to solve problem~(\ref{eq:form2}) due to the high dimensionality of audio. It is necessary to find low-dimensional perturbations and upsample to obtain the adversarial perturbations. However, upsampling techniques commonly used in vision, such as linear, bilinear, and nearest neighbor interpolations, are not suitable for audio because they produce a poor reconstruction of the signal, especially when a substantial reduction in the feature space is needed. Another idea suggested by Guo et al.~\cite{guo2019lowfreq} is to use the low-frequency spectrum of the signal. However, we found that a large number of frequency components should be perturbed to achieve any change in the transcription output, making it an inefficient basis for Bayesian optimization. Thus, we generate a random basis and use it as a linear map to transform a low-dimensional perturbation direction to the original dimension. According to our studies, this simple idea performs better than previous techniques. In our experiments, we set the number of basis vectors to be 100.

\subsection{Experimental results}
\begin{table}
  \caption{Performance of black-box attack methods given a budget of 5,000 queries. Mean (and standard deviation) across five different runs.}
  \label{tab:results}
  \centering
  \begin{tabular}{l c c}
    \toprule
    Method & $\mathbf{\ell_{\infty}}\downarrow$ & SNR$\uparrow$\\
    \midrule
    BayesOpt & 0.064 (0.001) & 19.840 (0.159) \\
    SignOpt & 0.028 (0.001) & 29.315 (0.417) \\
    NP-Attack & \textbf{0.022 (0.000)} & \textbf{29.410 (0.298)}\\
    \bottomrule
  \end{tabular}
\end{table}

In the first experiment, we compare the quality of the AEs generated with a fixed query budget of 5,000 queries. Each attack method is executed five times under different random seeds. The results are shown in Table~\ref{tab:results}. On average, NP-Attack manages to find audio AEs with the best quality as it yields the lowest mean $\ell_{\infty}$-norm and the highest average SNR. SignOpt has a larger variance because its performance relies on a good initialization. Interestingly, although NP-Attack can be seen as a variant of Bayesian optimization, it outperforms BayesOpt by a large margin. The reason could be that our neural predictor can capture the interactions in audio, which are highly structured. In contrast, the performances of Bayesian methods heavily rely on the choice of kernel functions.

In the second experiment, we demonstrate the efficiency of NP-Attack compared to other black-box methods. Figure~\ref{fig:success} shows the average success rate against the number of queries for different perturbation budgets $\lambda_{\text{max}}$. NP-Attack requires the least number of queries to achieve a high success rate. Importantly, it converges faster to the high success rates than the other approaches. For a large budget $\lambda_{\text{max}}=0.1$, only roughly 100 queries are sufficient for a successful attack with NP-Attack.

\begin{figure}[t]
    \centering
    \includegraphics[width=0.90\linewidth]{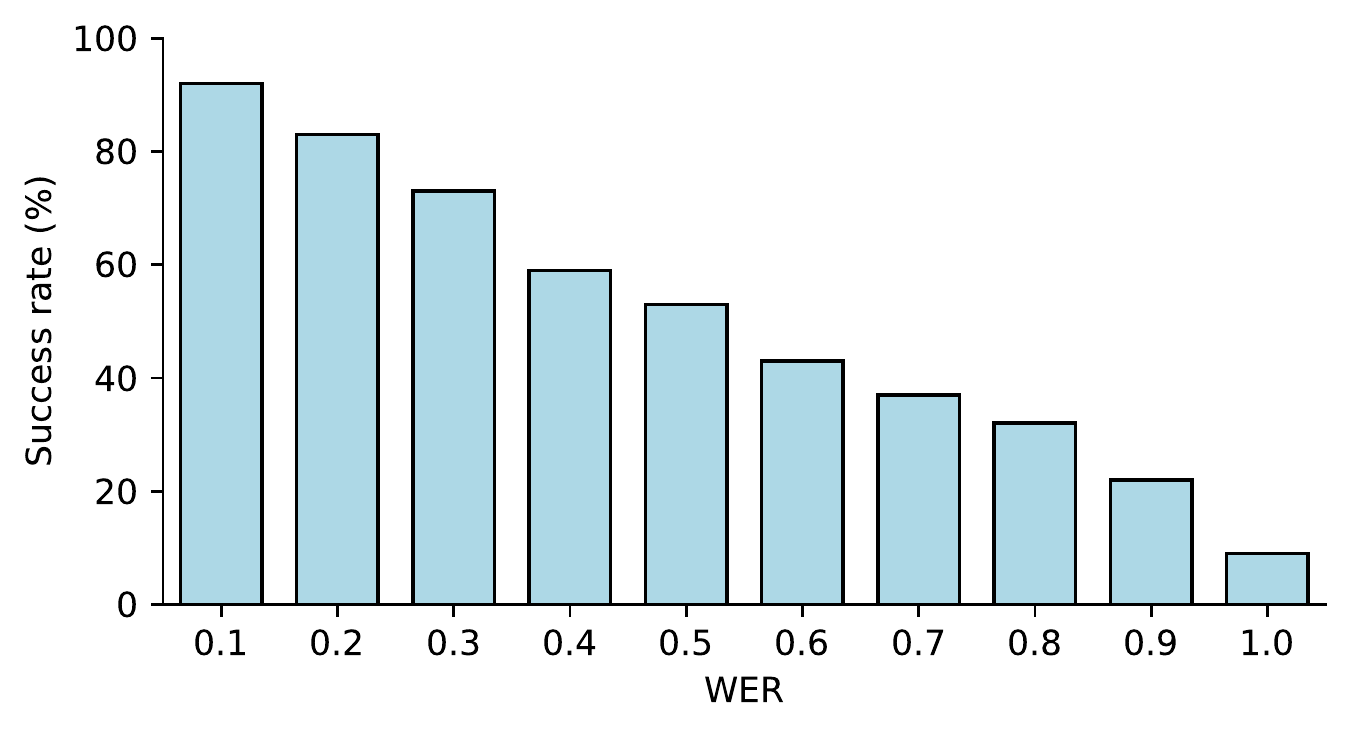}
    \vspace{-0.3cm}
    \caption{Success rate of NP-Attack for different minimum required WER between the original and adversarial transcript.}
    \label{fig:barplot}
    \vspace{-0.3cm}
\end{figure}

Finally, we conduct an ablation study to see the effect of varying the required WER on the success rate of NP-Attack. An attack is considered a success when the WER between the transcript of the AE and that of the original example is satisfied. Here, we use a perturbation budget of $\lambda_{\text{max}} = 0.05$ and a query budget of 5,000. The results are shown in Fig.~\ref{fig:barplot}.  As expected, by increasing the minimum WER, the proportion of successful attacks decreases since the more changes to the transcript we require, the more challenging the problem becomes. This is because the perturbation needs to fool the ASR model even in the locations where the model has highly confident predictions.

\section{Conclusions and future work}
In this paper, we have introduced NP-Attack, a novel predictor-based method to generate audio AEs on black-box ASR systems with high query efficiency. The proposed method leverages a neural predictor, which estimates the distance to the decision boundary for a given perturbation direction. We demonstrated that NP-Attack achieves high success rates with fewer queries while producing AEs that are close to the original.

Studying untargeted attacks can already be beneficial for building robust ASR systems. In future work, we will further extend our method to targeted attacks. It would also be interesting to study the performance of NP-Attack on different network architectures of the neural predictor.

\bibliographystyle{IEEEtran}

\bibliography{mybib}

\end{document}